\documentclass[11pt]{article}

\usepackage{epsfig}

\begin{document}

\title{
\mbox{}	\\
[-3cm]
{\footnotesize \hspace* {\fill} CFNUL - 99/03} \\
[2cm]
{\bf Velocity at the Schwarzschild horizon revisited}}

\author{ Ismael Tereno \\
{\em Centro de Fisica Nuclear, Universidade de Lisboa} \\
{\em 1649-003 Lisboa, Portugal}
}
\date{}
\maketitle

\begin{abstract}
The question of the physical reality of the black hole interior
is a recurrent one. An objection to its existence is the well
known fact that the velocity of a material particle, refered to the
stationary frame, tends to the velocity of light as it approaches
the horizon.

It is shown, using Kruskal coordinates, that a timelike radial geodesic
does not become null at the event horizon.
\end{abstract}

The interpretation of the maximal analytic extension of the 4 regions
Schwarzschild spacetime presents some difficulties. The conventional
view is that the only regions relevant to a black hole formed by
gravitational collapse are regions I and II \cite{luminet}.

There is however a long list of literature where the physical reality
of the black hole interior (region II) is argued. References can be
found elsewhere \cite{cooper}.

Recently a new case was made to that point \cite{mitra}. There, the
reasoning is mainly based in the result that the velocity of any material
particle as measured by a Kruskal observer (as defined below) is equal to
1 at the event
horizon. The purpose of this paper is to show that this is not the case.

\vspace {2ex}

In Schwarzschild coordinates, the metric of the Schwarzschild spacetime
takes the well known form,

\begin{equation}
ds^2=-\left( {1-{{2m} \over r}} \right)dt^2+\left( {1-{{2m} \over r}}
\right)^{-1}dr^2+r^2\left( {d\theta^2 +\sin^2 \theta d\varphi ^2} \right).
\label{metschw}
\end{equation}

For $r>2m$, the Kruskal coordinates $(x',t')$ relate to these by,

\begin{equation}
\label{xtext}
\left\{
\begin{array}{lll}
\mbox{$x'^2-t'^2=\left( {{{r-2m} \over {2m}}} \right)e^{r/2m}$} \\ \\
\mbox{$t'=\tanh \left( {{t \over {4m}}} \right)x'$}
\end{array}
\right.
\end{equation}
In these coordinates the metric takes the form \cite{mtw},

\begin{equation}
\label{kruskal}
ds^2={{32m^3}\over {r}}e^{-r/2m}(-dt'^2+dx'^2)+r^2\left(
{d\theta ^2+\sin ^2\theta d\varphi ^2} \right).
\end{equation}

A Kruskal observer is one which maintains the space-like coordinate $x'$
constant and consequentely, from (\ref{kruskal}), it verifies,

\begin{equation}
\label{movkru}
{{32m^3} \over {re^{r/2m}}}\left( {{{dt'} \over {d\tau }}} \right)^2=1.
\end{equation}
Differentiating (\ref{xtext}) we get,

\begin{eqnarray}
\label{ddtau}
{{{dr} \over {d\tau }}}={{8m^2} \over {e^{r/2m}r}} \left( x'{{{dx'} \over
{d\tau }}} -t' {{{dt'} \over {d\tau }}}\right), & &
{{dt} \over {d\tau }}=\left( {x'{{dt'} \over {d\tau }}-t'{{dx'} \over
{d\tau }}} \right){{8m^2} \over {e^{r/2m}(r-2m)}}.
\end{eqnarray}
Using $dx'=0$ and (\ref{movkru}) we can write the following equation :

\begin{equation}
\left( {1-{{2m} \over r}} \right)\left( {{{dt} \over {d\tau }}} \right)^2
-\left( {1-{{2m} \over r}} \right)^{-1}\left( {{{dr} \over {d\tau }}}
\right)^2={{2m(x^2-t^2)} \over {e^{r/2m}(r-2m)}}=1,
\end{equation}
meaning the Kruskal observer follows a radial trajectory.

Consider now a material particle along a radial ingoing trajectory in
region I. Its velocity, measured by a Kruskal observer is simply

\begin{equation}
v={{dx'}\over{dt'}},
\end{equation}
since (\ref{kruskal}) is diagonal with $g_{x'x'}=-g_{t't'}$. Dividing one
of the equations (\ref{ddtau}) by the other and solving for $v$
we obtain,

\begin{equation}
\label{v}
v={{1+t'/x'{{dt}\over{dr}}(1-2m/r)}
\over{t'/x'+{{dt}\over{dr}}(1-2m/r)}},
\end{equation}
which is the equation (20) of \cite{mitra}.

At the event horizon, separating regions I and II, the coordinates take
the values: $r=2m$, $t=+\infty$, $t'/x'=\tanh (t/4m)=1$ and $x' \ne 0$.
Apparently, making $t'/x'=1$ shows that when the particle, following any
trajectory described by $dt/dr$, and the observer intersects at the
horizon, the velocity measured by the latter is 1. However, $v$ is a
function of 2 coordinates ($r$ and $t$), and both limits must be taken
simultaneously.

To illustrate this point, consider the movement of a particle $p$ in
special relativity, relative to 2 referencials $S$ and $S'$ with
all the movements parallel to each other. The well known expression
for the addition of velocities is,

\begin{equation}
v'_{p/S'}=\left({{v_{p/S}-V_{S'/S}}\over{1-v_{p/S}V_{S'/S}}}\right).
\end{equation}
If both particle and $S'$ are moving at the speed of light with respect
to $S$, their relative velocity is not necessarily 1, the expression
giving ${0 \over 0}$. However, if we take only the limit $v_{p/S}=1$
we obtain identical expressions in the numerator and denominator.

So at this point we cannot determine the value of $v$ in (\ref{v}) in
general. Let us assume a specific trajectory $dt/dr$: a geodesic.
In this case there is a conserved quantity for motion \cite{mtw},

\begin{equation}
E={{dt} \over {d \tau}} \left(1-{{2m} \over {r}}\right).
\end{equation}
Inserting this into (\ref{metschw}) we get, for the ingoing geodesic,

\begin{equation}
{{dt} \over {dr}}=-E\left(1-{{2m} \over {r}}\right)^{-1}
\left[ E^2-\left(1-{{2m} \over {r}}\right) \right]^{-1/2}
\end{equation}
and (\ref{v}) becomes,

\begin{equation}
\label{v2}
v={{\sqrt{E^2-(1-2m/r)}-E \tanh (t/4m)}
\over{\sqrt{E^2-(1-2m/r)}\tanh (t/4m)-E}}.
\end{equation}
If we Taylor expand the square root in the vicinity of the horizon we
obtain,

\begin{equation}
\label{v3}
v={{1-\tanh (t/4m)-(r-2m)/(2rE^2)}
\over{\tanh (t/4m)-1-(r-2m)/(2rE^2)\tanh (t/4m)}} \to
-{{\varepsilon - \delta /E^2}\over {\varepsilon + \tanh(t/4m)
\delta /E^2}},
\end{equation}
where,
\begin{eqnarray}
\varepsilon = 1-\tanh(t/4m), & & \delta =(r-2m)/2r.
\end{eqnarray}
In this form we see that in the denominator there is a sum with
$\varepsilon$ while in the numerator there is a subtraction
from the same factor. We conclude that the modulus of $v$ is less than 1.

This expression also shows that, depending on the relative size of
$\varepsilon$ and $\delta$, $dx'/dt'$ can be negative or positive,
unlike $dr/dt$ which is obviously always negative in an ingoing
geodesic. For null geodesics where $E = \infty$, we must obtain
$v=-\varepsilon / \varepsilon = -1$.

This discussion is reminiscent of an equivalent one that took place
over 20 years ago \cite{cavspin1,janis1,cavspin2,janis2}. In that
case the problem was not posed in terms of Kruskal coordinates but was
solved with the use of another set of ingoing coordinates. It is an
example of a different observer who measures a sub-luminous velocity
at the event horizon.

\begin{figure}[ht]
\begin{center}
\epsfig{file=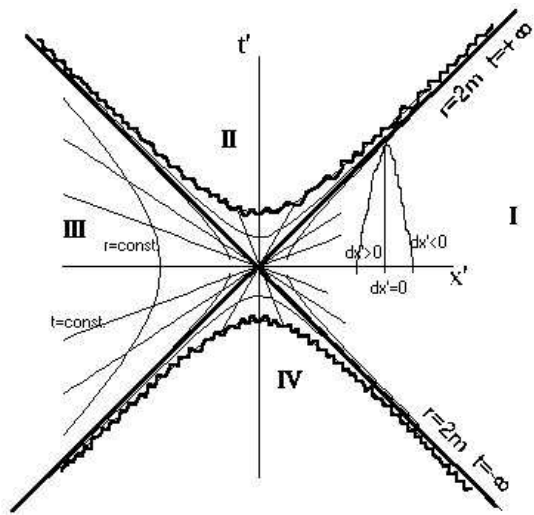, height=7cm}
\caption{Kruskal diagram.}
\end{center}
\end{figure}

\end{document}